\pdfoutput=1
\documentclass[sigconf]{acmart}
\usepackage{listings}
\usepackage{algorithm}
\usepackage{algorithmic}

\copyrightyear{2026}
\acmYear{2026}
\setcopyright{cc}
\setcctype{by}
\acmConference[SEAMS '26]{21st International Conference on Software Engineering for Adaptive and Self-Managing Systems}{April 13--14, 2026}{Rio de Janeiro, Brazil}
\acmBooktitle{21st International Conference on Software Engineering for Adaptive and Self-Managing Systems (SEAMS '26), April 13--14, 2026, Rio de Janeiro, Brazil}
\acmPrice{}
\acmDOI{10.1145/3788550.3794870}
\acmISBN{979-8-4007-2445-9/2026/04}

\begin{document}

\title[Software Self-Extension with \textsc{SelfEvolve}]{Software Self-Extension with \textsc{SelfEvolve}: an Agentic Architecture for Runtime Code Generation}

\author{Md Asif Iqbal Fahim}
\affiliation{\institution{University College Dublin (UCD)}\city{Dublin}\country{Ireland}}
\email{md.fahim@ucdconnect.ie}

\author{Oluwadamilola Adebayo}
\affiliation{\institution{JP Morgan Chase}\city{Dublin}\country{Ireland}}
\email{dami.adebayo@jpmchase.com}

\author{Alessio Ferrari}
\affiliation{\institution{University College Dublin (UCD)}\city{Dublin}\country{Ireland}}
\email{alessio.ferrari@ucd.ie}

\renewcommand{\shortauthors}{Fahim, Adebayo, and Ferrari}

\begin{abstract}
Traditional self-adaptive systems automatically reconfigure existing components in response to changing requirements, but provide limited support for the generation of novel functionalities. The software generation capabilities of large language models (LLMs) open the possibility to create entirely new modules at runtime, enabling a form of self-evolution beyond traditional self-adaptation. 
We present \textsc{SelfEvolve}, an orchestrated agentic pipeline architecture enabling runtime self-extension---the autonomous addition of new capabilities during execution---as a preliminary form of self-evolution.  Self-extension focuses on the autonomous generation and integration of new functions, based on user requests, without requiring a system restart or developer intervention. Evaluation of our architecture across 11 self-extension tasks demonstrates an average Pass@1 of 92.7\% (51/55), outperforming developer-focused code generation baselines like AutoGen, MetaGPT, and AgentCoder. \textsc{SelfEvolve} achieves 61.8\% improvement over the best baseline, i.e. Autogen, with statistical significance.  This work demonstrates the feasibility of runtime capability extension through autonomous code generation. This provides preliminary evidence for a paradigm in which  systems autonomously evolve to satisfy user needs, paving the way towards individualised, self-improving systems.
\end{abstract}


\maketitle

\section{Introduction}
Software systems typically operate with fixed capabilities determined at development time. Adding new functionality requires human developers to write code, test it, and deploy updates---a process taking days to weeks \cite{dora2024}. Automated code generation with LLM agents~\cite{dong2025survey} offers the possibility to accelerate development time, with speed gains of about 21\%, according to a controlled experiment at Google~\cite{paradis2025much}.
Commercial and open source code generation tools include GitHub Copilot~\cite{copilot}, CodeT5~\cite{wang2021codet5}, and TiCoder~\cite{ticoder}. The scientific literature also offers several solutions, such as ClarifyGPT~\cite{mu2024clarifygpt}, empowering code generation with requirements clarification, the self-planning approach by Jiang et al.~\cite{jiang2024self} for automatic problem decomposition, and multi-agent frameworks like MetaGPT~\cite{hong2023metagpt},  AgentCoder~\cite{huang2024agentcoder}, and AutoGen~\cite{wu2023autogen}. These systems accelerate \textit{development-time} code generation based on technical requirements specified by developers, producing code for human developers to integrate offline. In contrast, self-evolution through \textit{runtime} code generation---where systems autonomously extend their own capabilities based on user needs expressed as high-level requirements---remains unexplored. Consider a Netflix user requesting ``show me what people in my city are watching.'' Rather than returning an error for missing neighborhood analytics, the system autonomously generates this capability---synthesizing, validating, and integrating the function into its runtime, then transparently delivering results. This vision of self-evolution is oriented to enable \textit{individualized software} for consumer applications---systems where each user has uniquely generated capabilities tailored to their needs, with the user as the only human on the loop, removing the developer and minimizing their role. Over time, commonly requested functionalities can be applied to other users requesting similar capabilities, enabling both individualization and targeted diffusion across user groups. 

Attempts at self-evolution exist in the literature. Multi-Agent Self-Evolving (MASE) systems~\cite{fang2025comprehensivesurveyselfevolvingai} typically refine prompts, memory, and tool-use strategies (e.g., reaching services through MCP\footnote{\url{https://modelcontextprotocol.io/}}) but do not generate new functions during operation. Evolutionary computation studies~\cite{wu2024evolutionary} using genetic algorithms have examined LLMs' potential to generate program variations~\cite{zhao2023survey}, yet these variants are produced in isolation without integration mechanisms. Recent LLM-based code refinement work~\cite{guo2024exploring,jin2025reveal} focuses on iterative improvement without considering runtime evolution.

In this paper, we present \textsc{SelfEvolve}, an architecture enabling runtime self-extension---the autonomous addition of new capabilities during execution---as a preliminary form of self-evolution. At this preliminary stage, we focus on generation of new functions that integrate with existing codebases, not structural modification or refactoring. This approach enables runtime capability acquisition without system restart. Our architecture consists of coordinating specialized components with defined interfaces---dispatcher, test generator, code synthesizer, sandboxed executors, and context memory---ensuring isolation between generation and core functionality of the software to evolve. The architecture embeds test-driven development (TDD)~\cite{beck2002test}, a methodology where test cases are generated before code to guide correct implementation. The system implements iterative refinement---repeated improvement cycles that progressively correct failures---through dual feedback mechanisms: a two-stage validation process using both initial TDD tests and comprehensive unit tests. When generated code fails validation, adjudicator components analyze test execution results to produce feedback that constrains subsequent generation attempts. This feedback mechanism accumulates constraints across iterations, progressively narrowing the solution space until the generated code passes all test cases or reaches the iteration limit. Our evaluation addresses two research questions (RQs): 

\textbf{RQ1:} \textit{What is the performance of \textsc{SelfEvolve} for runtime  self-extension across diverse software tasks?} We evaluate the framework on 11 problems categorized into integration (4 problems, 577-783 LOC), compositional (3 problems, 30-50 LOC), and data processing (4 problems, 50-100 data records) tasks. We also perform quantitative baseline comparison against AgentCoder, AutoGen, and MetaGPT on the same tasks. The framework achieves 92.7\% average Pass@1~\cite{chen2021evaluating,austin2021program,yin2022natural} (51/55 successful runs), outperforming the best baseline AutoGen by 61.8\% improvement, with 2.2 average iterations to convergence. 

\textbf{RQ2:} \textit{To what extent does the TDD pipeline contribute to  self-extension performance?} We conduct an ablation study comparing framework configurations with and without the TDD pipeline. The baseline configuration omits the intermediate adjudication phase, proceeding directly from function generation to final validation, enabling quantification of the TDD pipeline's impact on both success rates and convergence characteristics. Results demonstrate statistically significant improvements with TDD: TDD configuration achieves 92.7\% Pass@1 versus 72.7\% without TDD, with 2.1× faster convergence (2.2 vs 4.7 average iterations).

\textbf{Data Availability.} Implementation: \url{https://anonymous.4open.science/r/self-evolving-systems-FEC4}. Replication Package: \url{https://zenodo.org/records/18152531}.
\nopagebreak
\section{Comparison with Other Agentic Frameworks}
Multi-agent frameworks for LLM-based code generation employ iterative refinement with validation loops, making them most comparable to \textsc{SelfEvolve}. Among these agentic architectures, we compare with three representative state-of-the-art systems: AgentCoder~\cite{huang2024agentcoder} uses multi-agent collaboration for test-driven generation; AutoGen~\cite{wu2023autogen} provides conversation-based infrastructure for 
 agent interactions; MetaGPT~\cite{hong2023metagpt} employs role-based agents simulating development teams. Table~\ref{tab:architecture-comparison} presents a qualitative comparison. Detailed capability definitions are provided in the replication package. The comparison was performed through systematic analysis of original papers (see replication package for details, including quotations of the papers). Overall, existing frameworks are general-purpose for development-time code generation; \textsc{SelfEvolve} is a specialized architecture for runtime self-extension. Existing frameworks are developer oriented and offer offline code generation, while \textsc{SelfEvolve} enables user-centered runtime self-extension. All platforms share code generation capabilities (multi-agent collaboration, iterative refinement, test-based validation). The main differences are: context management (automatic codebase analysis, multi-file navigation, import resolution), persistence mechanisms (persistent knowledge base, automated code reuse), and runtime system integration (hot code reloading). A strong distinction is user orientation---existing frameworks are designed to respond to technical requirements from developers, while \textsc{SelfEvolve} responds to requests from end users. This is enabled by specialized high-to-low level requirements transformation (via a Chat \& Tool Dispatcher component) and runtime extension facilitating fast user validation. Effectiveness is confirmed by our evaluation, where user requirements are given as input.
 \begin{table}[!htbp]
\centering
\setlength{\tabcolsep}{2pt}
\caption{Architectural comparison for runtime self-extension}
\label{tab:architecture-comparison}
\footnotesize
\begin{tabular}{lcccc}
\toprule
Capability & Agent\-Coder & Auto\-Gen & Meta\-GPT & \textbf{Self\-Evolve} \\
\midrule
\textit{Code Generation} & & & & \\
\quad Multi-agent collaboration & Yes & Yes & Yes & \textbf{Yes} \\
\quad Iterative refinement & Yes & Yes & Yes & \textbf{Yes} \\
\quad Test-based validation & Yes & Yes & Yes & \textbf{Yes} \\
\midrule
\textit{Context Management} & & & & \\
\quad Automatic codebase analysis & No & Partial & No & \textbf{Yes} \\
\quad Multi-file navigation & No & Partial & Partial & \textbf{Yes} \\
\quad Import path resolution & No & No & Partial & \textbf{Yes} \\
\midrule
\textit{Persistence \& Reuse} & & & & \\
\quad Persistent knowledge base & No & No & No & \textbf{Yes} \\
\quad Automated code reuse & No & No & No & \textbf{Yes} \\
\midrule
\textit{System Integration} & & & & \\
\quad User-oriented & No & No & No & \textbf{Yes} \\
\quad Hot code reloading & No & No & No & \textbf{Yes} \\
\bottomrule
\end{tabular}
\Description{Comparison table showing capabilities across four frameworks: AgentCoder, AutoGen, MetaGPT, and SelfEvolve, organized into code generation, context management, persistence and reuse, and system integration categories.}
\end{table}
\nopagebreak
\section{\textsc{SelfEvolve} Architecture}
\textsc{SelfEvolve} implements an agentic architecture for autonomous runtime code generation, where each processing component operates as an LLM agent. This is an autonomous module that leverages an LLM to pursue goals by analyzing inputs, maintaining context, making decisions, and generating outputs. Three technical solutions enable runtime updates in \textsc{SelfEvolve}: persistent knowledge base---a component maintaining generated functions and their specifications for future reuse---dynamic module integration, and process isolation. 
\begin{figure*}[!t]
\centering
\includegraphics[width=0.9\textwidth]{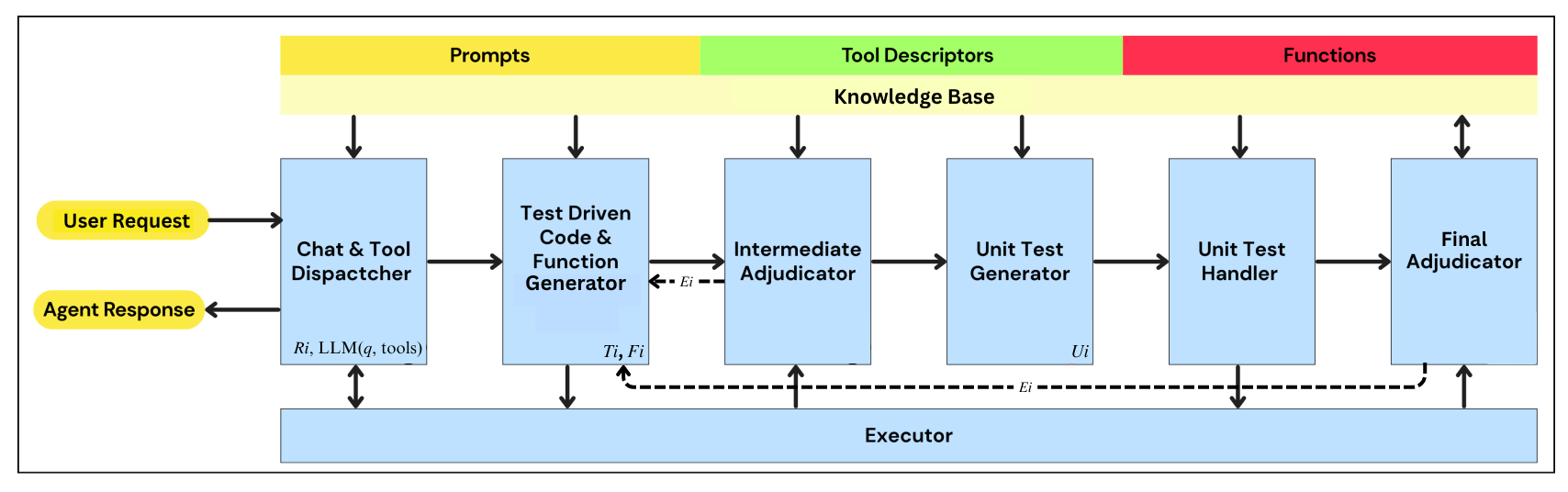}
\caption{Runtime code generation pipeline from user requests to generation,  verification, and integration.}
\Description{Diagram showing the test-driven generation pipeline with steps: TDD test generation, function synthesis, intermediate adjudication, unit
test generation, and final adjudication.}
\label{fig:architecture}
\end{figure*}
In the standard behavior, a user requests a functionality in natural language (using OpenAI function-calling format~\cite{openai2025}, this is already available in the knowledge base, and it is thus executed. The evolution process begins when users request an unavailable functionality: the system generates code and creates a separate subprocess sandbox for safe execution and testing. Following successful verification, the function is promoted from this sandbox to permanent system integration: added to the knowledge base and immediately accessible without restart (through the \texttt{importlib.reload()} Python feature), achieving runtime self-extension. In the following, we provide the details of our architecture.
\subsection{Orchestrated Pipeline Architecture}
Figure~\ref{fig:architecture} illustrates the architecture where each component is coordinated by deterministic control flow. 

The \textbf{Knowledge Base} (shown at the top of Figure~\ref{fig:architecture}) comprises three components. The \textit{Prompts} repository stores system instructions and few-shot examples~\cite{openai2024prompting} to be used by components, as well as additional problem-specific prompts generated throughout the process. The \textit{Functions} repository contains executable Python implementations. The \textit{Tool Descriptors} repository maintains JSON specifications that pair functions with their invocation metadata, enabling the OpenAI function-calling protocol---we refer to these function-descriptor pairs as \textit{Tools} to distinguish them from standalone \textit{Functions}. The system processes user requests through six sequential components, with the Knowledge Base providing context and the \textbf{Executor} layer ensuring safe execution. A key capability is that newly generated functions can utilize available resources (existing codebases, data files, or previously generated functions). The pipeline is iterative: the state at iteration $i$ consists of a user request $R_i$, TDD test cases $T_i$ based on $R_i$, function code $F_i$, comprehensive unit tests $U_i$ (e.g., for edge cases), and feedback $E_i$ (a single message aggregating all needed corrections from failed tests). In the following, $\text{LLM}(\cdot)$ denotes invocation of GPT-4.1 with component-specific prompts from the Prompts repository. While the base model is identical across components, the prompts are tailored for each specific component based on OpenAI prompt engineering guidelines~\cite{openai2024prompting}. Our evaluation focuses on the complete TDD pipeline as a unified architectural element, as test generation and intermediate adjudication function together in our design. We illustrate \textsc{SelfEvolve} through eigenvalue calculation: a user requests ``compute eigenvalues of this matrix'', but the function is not available. However, the Functions repository contains the \texttt{matrix\_operations} function (performing matrix multiplication), which eigenvalue calculation can use as a sub-function. 

\textbf{Chat \& Tool Dispatcher.} The entry component in Figure~\ref{fig:architecture} (an agent embedding an LLM) receives user requests and determines if existing tools from the Knowledge Base suffice, by searching available Tools. The LLM either returns an existing tool call $\text{tool\_call}(f_j)$ if matched, or produces requirement $R_i$ if a new tool is needed. Finding no eigenvalue calculation function, the dispatcher produces requirement $R_i$ by transforming the user's request into a structured specification, made of desired function name and requirements: < \texttt{compute\_eigenvalues}, ``compute eigenvalues for a given matrix input.''>. More formal constraints could also be generated, e.g., pre/post-conditions, but at this stage we kept the structure simple given the preliminary nature of the approach. 

\textbf{Test-Driven Code \& Function Generator.} This dual-purpose component first generates TDD test cases that define inputs and expected outputs: $T_i = \text{LLM}(R_i + E_{i-1})$; where $E_{i-1}$ contains reinforcement from previous iterations---feedback from previously failed attempts, e.g., ``Function returned incorrect eigenvalues: expected [2.0, 3.0] but received [1.7, 3.4]; numerical instability detected''. The test count is not predetermined; instead, the LLM autonomously generates tests guided by loosely specified prompts that encourage comprehensive coverage---future versions will include more formal structural metrics as coverage target. This results in variable test counts across problems. The component then generates the function implementation to satisfy these tests: $F_i = \text{LLM}(R_i, T_i)$. When generating the function, the LLM examines available resources in the Functions repository, identifies matrix multiplication capabilities, and generates a function \texttt{compute\_eigenvalues} that imports and calls \texttt{matrix\_operations}. If no function existed that could be leveraged, the component would have generated the entire code.

\textbf{Intermediate Adjudicator.} The pipeline employs two adjudicators, i.e., automated evaluators that analyze test execution results to make pass/fail decisions about generated code: Intermediate and Final Adjudicator. Here we describe the former. The Intermediate Adjudicator validates generated code against TDD tests, executing the function and verifying results: $\text{test\_output} = \text{execute\_tests}(F_i, T_i)$. It then evaluates results via $\text{LLM}(\text{test\_output})$: if all tests pass, it produces \textit{accept}; if tests fail, it produces \textit{reject} along with feedback $E_i$ specifying error phenomena. In case of \textit{reject}, control returns to the function generation component for another attempt. 

\textbf{Unit Test Generator.} After passing intermediate adjudication, the pipeline generates comprehensive unit tests $U_i = \text{LLM}(F_i, R_i)$ based on the actual generated code. Unlike TDD tests (which operate on requirements only), unit tests examine the specific implementation to validate error handling and edge cases specific to the generated solution. 

\textbf{Unit Test Handler.} The handler executes unit tests through the Executor layer, with 30-second timeouts preventing infinite loops. It captures stdout, stderr, exit codes, and returns test results for final adjudication. 

\textbf{Final Adjudicator.} The second adjudicator evaluates results via $\text{LLM}(\text{test\_results})$, producing \textit{accept} when all tests pass or \textit{reject} with feedback $E_i$ when tests fail. Acceptance adds the function to the Knowledge Base. Rejection triggers the feedback loop, where $E_i$ is provided with $R_i$ to the Test-Driven Code \& Function Generator for the next iteration, similarly to the Intermediate Adjudicator. The iteration process for both adjudicators (represented as dashed back arrows in Figure~\ref{fig:architecture}) continues until a maximum number of iterations $n$ is reached (computed as the sum of the cycles due to the two components). The value $n$ is set to 6 in our experiments.

\textbf{System Integration and Response.} Upon acceptance, the function permanently integrates into the system through three updates: the function code is written to the Functions repository, its JSON descriptor is added to the Tool Descriptors repository enabling future discovery, and the Prompts repository is updated with patterns recognizing when to invoke this function (e.g., ``When asked to produce eigenvalue, call \texttt{compute\_eigenvalues}''). This integration constitutes the system's next release---the function becomes a permanent capability immediately available to all future requests. Following integration, the Chat \& Tool Dispatcher invokes the newly integrated function with the user's original input. The Executor layer executes the function in a subprocess sandbox and returns the results. The Chat \& Tool Dispatcher then receives these results and generates a natural language response to the user, transparently fulfilling the request---for the eigenvalue example, computing and returning the eigenvalues as requested.
\nopagebreak
\section{Preliminary Evaluation and Discussion}
\label{sec:preliminary-evaluation}
\subsection{Evaluation Methodology}
\textbf{Dataset.} We manually created 11 problems including their codebases, datasets, and ground truth test specifications spanning 8 application domains (problems specified in Table~\ref{tab:problem-characteristics}). The dataset comprises three task categories: (1) \textit{Integration tasks} (4 problems) requiring analysis and interfacing with existing multi-file codebases (577-783 LOC, 7-20 classes) to generate functions that integrate with existing class hierarchies; (2) \textit{Compositional tasks} (3 problems) testing temporal composition where subsequently generated functions build upon previously created capabilities; (3) \textit{Data processing tasks} (4 problems) requiring manipulation of external datasets (50-100 records) with operations like filtering, aggregation, and recommendation. Our integration tasks (577-783 LOC) represent a step forward compared to standard code generation benchmarks: HumanEval and MBPP average 11.5 and 6.8 LOC respectively~\cite{du2024classeval}. Each problem includes predefined test specifications with assertions verifying behavioral correctness: type checks (e.g., \texttt{assert isinstance(result, dict)}), structural constraints (e.g., \texttt{assert `titles' in result}), value ranges (e.g., \texttt{assert 3.0 <= avg\_rating <= 5.0}), and edge cases. These ground truth test specifications were manually created by the authors, covering all described functionalities, edge cases, and output validation for specific test inputs. Completeness was ensured through peer review among all authors. 
\begin{table}[!t]
\centering
\setlength{\tabcolsep}{3pt}
\caption{ Self-extension problem characteristics}
\label{tab:problem-characteristics}
\footnotesize
\begin{tabular}{lccc}
\toprule
Problem & Domain & Data Description \\
\midrule
\textbf{Integration Tasks} & & & \\
\quad Salary Analyzer & HR Analytics & 616 LOC, 13 files, 12 classes \\
\quad Patient Risk Analyzer & Healthcare & 752 LOC, 12 files, 20 classes \\
\quad Student GPA Calculator & Education  & 783 LOC, 10 files, 10 classes \\
\quad Inventory Low Stock Alert & E-Commerce & 577 LOC, 8 files, 7 classes \\
\textbf{Compositional Tasks} & & & \\
\quad Matrix Eigenvalue & Linear Algebra & Prior func. (40-50 LOC) \\
\quad Portfolio Risk & Finance & Prior func. (30-40 LOC) \\
\quad IoT Sensor Pipeline & IoT Systems & Prior func. (40-50 LOC) \\
\textbf{Data Processing Tasks} & & & \\
\quad Movie API & Media Systems & 100 records \\
\quad Book Recommender & Content Rec. & 100 records \\
\quad Performance Tracker & HR Analytics & 50 records \\
\quad Friend Suggester & Social Network & 100 users \\
\bottomrule
\end{tabular}
\Description{Table listing 11 self-extension problems with their domains, task types, and data characteristics including lines of code, file counts, and record counts.}
\end{table}
\begin{table}[!htbp]
\centering
\setlength{\tabcolsep}{.5pt}
\caption{Framework comparison (Avg. Pass@1, 5 runs)}
\label{tab:rq1-results}
\footnotesize
\begin{tabular}{lcccc}
\toprule
Problem & AgentCoder & AutoGen & MetaGPT & \textbf{\textsc{SelfEvolve}} \\
\midrule
\textbf{Integration Tasks} & & & & \\
\quad Salary Analyzer & 0/5 (0\%) & 0/5 (0\%) & 2/5 (40\%) & \textbf{4/5 (80\%)} \\
\quad Patient Risk Analyzer & 0/5 (0\%) & 0/5 (0\%) & 0/5 (0\%) & \textbf{5/5 (100\%)} \\
\quad Student GPA Calculator & 0/5 (0\%) & 0/5 (0\%) & 5/5 (100\%) & \textbf{5/5 (100\%)} \\
\quad Inventory Low Stock Alert & 0/5 (0\%) & 0/5 (0\%) & 3/5 (60\%) & \textbf{4/5 (80\%)} \\
\quad \textit{Subtotal} & \textit{0/20 (0\%)} & \textit{0/20 (0\%)} & \textit{10/20 (50\%)} & \textit{\textbf{18/20 (90\%)}} \\
\midrule
\textbf{Compositional Tasks} & & & & \\
\quad Matrix Eigenvalue & 0/5 (0\%) & 1/5 (20\%) & 0/5 (0\%) & \textbf{4/5 (80\%)} \\
\quad Portfolio Risk & 0/5 (0\%) & 5/5 (100\%) & 4/5 (80\%) & \textbf{5/5 (100\%)} \\
\quad IoT Sensor Pipeline & 0/5 (0\%) & 5/5 (100\%) & 1/5 (20\%) & \textbf{5/5 (100\%)} \\
\quad \textit{Subtotal} & \textit{0/15 (0\%)} & \textit{11/15 (73.3\%)} & \textit{5/15 (33.3\%)} & \textit{\textbf{14/15 (93.3\%)}} \\
\midrule
\textbf{Data Processing Tasks} & & & & \\
\quad Movie API & 0/5 (0\%) & 0/5 (0\%) & 0/5 (0\%) & \textbf{4/5 (80\%)} \\
\quad Book Recommender & 0/5 (0\%) & 0/5 (0\%) & 0/5 (0\%) & \textbf{5/5 (100\%)} \\
\quad Performance Tracker & 0/5 (0\%) & 3/5 (60\%) & 0/5 (0\%) & \textbf{5/5 (100\%)} \\
\quad Friend Suggester & 0/5 (0\%) & 3/5 (60\%) & 0/5 (0\%) & \textbf{5/5 (100\%)} \\
\quad \textit{Subtotal} & \textit{0/20 (0\%)} & \textit{6/20 (30\%)} & \textit{0/20 (0\%)} & \textit{\textbf{19/20 (95\%)}} \\
\midrule
\textbf{Overall} & \textbf{0/55 (0\%)} & \textbf{17/55 (30.9\%)} & \textbf{15/55 (27.3\%)} & \textbf{51/55 (92.7\%)} \\
\bottomrule
\end{tabular}
\Description{Results table comparing Pass@1 success rates across AgentCoder, AutoGen, MetaGPT, and SelfEvolve frameworks for integration, compositional, and data processing tasks.}
\end{table}

\textbf{RQ1: Self-Extension Performance.}
We evaluate \textsc{SelfEvolve}'s performance across all 11 problems, executing each problem 5 times to account for output variability (55 total runs). We also perform quantitative baseline comparison with AgentCoder, AutoGen, and MetaGPT. For each problem, we computed the success proportion across its 5 runs (e.g., if 4 out of 5 runs succeeded, proportion = 0.80). We then applied pairwise Wilcoxon signed-rank tests to these 11 pairs of proportions (N=11 problem-level pairs, not N=55 runs) with Bonferroni correction for three comparisons ($\alpha$=0.05/3=0.0167), following standard practices~\cite{Malhotra2017, demsar2006}. Pairing at the problem level accounts for problem-specific difficulty while treating the 11 problems as independent observations. 

\textbf{RQ2: TDD Pipeline Impact (Ablation Study).}
We systematically compare performance with and without the TDD pipeline across all 11 problems, conducting 5 runs per problem in each configuration (110 total experiments: 55 with TDD, 55 without TDD). Using the same pairing methodology as RQ1, we applied Wilcoxon signed-rank test (Bonferroni correction not required, as only two treatments are considered). 

\textbf{Model Selection.} We conduct all experiments using GPT-4.1, selected specifically for its support of the OpenAI function-calling protocol, which is essential for our tool invocation mechanism. This protocol enables the system to map natural language requests to specific function calls with structured parameters, a capability not universally available across LLMs. 

\textbf{Evaluation Approach and Metrics.} Results are evaluated through automated pytest execution against ground truth test specifications included in the dataset. Generated functions must pass all pytest assertions to be considered successful. Success is binary: if the test exits with code 0 (all assertions pass), Pass@1 = 100\% for that run; otherwise Pass@1 = 0\%. Our ground truth is behavioral (test specifications), allowing multiple correct code implementations to satisfy the same requirements. All pytest execution results are manually double-checked by the authors to ensure validity. Following established practices~\cite{chen2021codex,chen2023selfdebug,madaan2023selfrefine}, we measure: (1) \textit{Average Pass@1}---the average Pass@1 across the 5 runs per problem; (2) \textit{Iteration Count}---average number of refinement cycles needed, indicating convergence efficiency.
\subsection{Results and Future Plans}
\textbf{RQ1:  Self-Extension Performance.}
Table~\ref{tab:rq1-results} presents the results. \textsc{SelfEvolve} achieved 92.7\% overall Pass@1 (51/55), outperforming the best baseline AutoGen by 61.8\% (92.7\% vs 30.9\%), with AgentCoder at 0.0\% and MetaGPT at 27.3\%. All comparisons are statistically significant: adjusted p-values (p$_{adj}$)=0.003, 0.012, 0.006 respectively. Iteration efficiency (average 2.2 iterations) is shown in Table~\ref{tab:rq2-tdd-comparison}, which presents results for the complete architecture with the TDD pipeline. Detailed quantitative analysis is provided in the replication package. 

\textbf{RQ2: TDD Pipeline Impact (Ablation Study).}
Table~\ref{tab:rq2-tdd-comparison} presents the TDD ablation study. TDD achieved statistically significant improvements over the baseline (Wilcoxon signed-rank test: W=55, p<0.001, r=0.98). Beyond 20\% success improvement (92.7\% vs 72.7\%), TDD enabled 2.1× faster convergence (2.2 vs 4.7 average iterations). Integration tasks showed largest gains (65\% to 90\%, 5.0 to 2.4 average iterations). Compositional tasks improved from 73.3\% to 93.3\% (4.7 to 2.1 average iterations). Data processing increased from 80\% to 95\% (4.4 to 2.2 average iterations). The large effect size with high significance demonstrates TDD provides both higher success rates and faster convergence.
\begin{table}[h]
\centering
\setlength{\tabcolsep}{2pt}
\caption{TDD impact (Avg. Pass@1, 5 runs)}
\label{tab:rq2-tdd-comparison}
\footnotesize
\begin{tabular}{lcccc}
\toprule
Task Category & \multicolumn{2}{c}{With TDD} & \multicolumn{2}{c}{Without TDD} \\
\cmidrule(lr){2-3} \cmidrule(lr){4-5}
& Pass@1 & Avg Iter & Pass@1 & Avg Iter \\
\midrule
Integration Tasks & 18/20 (90\%) & 2.4 & 13/20 (65\%) & 5.0 \\
Compositional Tasks & 14/15 (93.3\%) & 2.1 & 11/15 (73.3\%) & 4.7 \\
Data Processing Tasks & 19/20 (95\%) & 2.2 & 16/20 (80\%) & 4.4 \\
\midrule
\textbf{Overall} & \textbf{51/55 (92.7\%)} & \textbf{2.2} & \textbf{40/55 (72.7\%)} & \textbf{4.7} \\
\bottomrule
\end{tabular}
\Description{Ablation study table comparing TDD and non-TDD configurations showing Pass@1 rates and average iteration counts for each task category.}
\end{table} 

\textbf{Threats to Validity.} \textit{Internal Validity:}  (1) The dataset was manually handcrafted, limiting the diversity of problem types and complexity levels to those anticipated by the authors. (2) Ground truth tests might not be comprehensive. To mitigate this, we manually reviewed the generated code. (3) Using the same LLM for test and code generation creates test-code collusion risk. We mitigate this through modular architecture with separate agents using distinct system prompts and fresh API calls, following established practices~\cite{huang2024agentcoder,dong2024selfcollaboration}. Evaluation on novel code not used to train the LLMs also reduces this risk. Model diversity with multiple LLMs will further mitigate this in future work. (4) Integration of a new function does not guarantee that other functionalities remain correct. However, tasks were deliberately kept independent to isolate self-extension mechanics; therefore, the need for regression testing is limited in this preliminary evaluation. Comprehensive regression testing for interdependent function ecosystems is future work. \textit{External Validity:} (1) Single-function generation may not reveal multi-file generation challenges, and results from 11 problems may not generalize. Expanding the problem set and long-term deployment studies will address generalizability in future work. Considering domain applicability of the whole approach, we are oriented toward consumer software rather than safety-critical systems. (2) We used only GPT-4.1, performance may vary with other LLMs. Our architecture requires OpenAI function-calling protocol, limiting applicability to LLMs lacking this capability (e.g., open-source models).
\textit{Construct Validity:}  (1) Our evaluation focuses exclusively on functional correctness, deliberately isolating from performance, scalability, maintainability, security, and non-functional requirements in general. Non-functional requirements (security, scalability, performance, maintainability), along with code quality evaluation through smell identification~\cite{Femmer2014RequirementsSmells} and human evaluation, represent future work.
\textit{Conclusion Validity:} (1) Limited sample size (N=11 task-level pairs) constrains statistical power, which can lead to Type II errors. However, both RQ1 and RQ2 show statistical significance with large differences. More rigorous evaluation with increasing the number of problems for greater statistical power and comprehensive statistical analysis including power analysis will strengthen claims in future work. 

\textbf{Implications.} \textsc{SelfEvolve} transforms software development by enabling runtime capability acquisition without developer intervention. For research, demonstrating that systems can modify executing code opens directions in self-improving architectures, autonomous debugging, and evolutionary design. For practice, runtime self-evolution enables fast adaptation to user needs and rapid feedback loops, eliminating deployment bottlenecks while enabling mass personalization. 

\textbf{Future Directions.} Critical extensions include:
(1) \textit{MAPE-K loop integration}---implementing a complete MAPE-K \cite{mape-k} (Monitor-Analyze-Plan-Execute-Knowledge) loop typical of self-adaptive systems. While \textsc{SelfEvolve} currently operates reactively to user requests, a full MAPE architecture would enable \textit{proactive} self-evolution through continuous monitoring of generated function performance and usage patterns, analysis of execution metrics to identify optimization opportunities or recurring failure modes, planning of autonomous improvements without explicit user requests, and execution of self-initiated adaptations~\cite{dellarocas1998architecture,weyns2023vision}. This would transform the system from request-driven evolution to autonomous self-improvement based on runtime observations; (2) \textit{Human-on-the-loop and Guardrails}: besides verification of correct implementation, it is critical to validate the generated behavior with the user. Furthermore, in mission-critical contexts, guardrails~\cite{dong2025safeguarding} need to be introduced to prevent unwanted evolutions. Controlled experiments will be conducted to ensure that autonomous adaptations are interpretable, trustworthy, and consistent with user expectations, reinforcing human confidence in the system's self-evolution process. (3) \textit{Embed Requirements Engineering}: since our approach is user-oriented, we will focus especially on requirements engineering extensions to clarify user intent, understand context, and capture domain knowledge. This includes multi-agent debate~\cite{chan2023chateval}, requirements clarification~\cite{mu2024clarifygpt}, improved requirements formalization (e.g., pre/post-conditions), and change-impact analysis to improve robustness and scalability. (4) \textit{Runtime Instrumentation}: incorporate deeper runtime data (breakpoints, memory states, call stacks) to enhance failure diagnosis precision.
\bibliographystyle{ACM-Reference-Format}
\bibliography{references_arxiv}

@techreport{dora2024,
  title={{2024 Accelerate State of DevOps Report}},
  author={{DORA}},
  year={2024},
  institution={DevOps Research and Assessment},
  type={Technical Report},
  url={https://dora.dev/research/2024/dora-report/},
  note={Accessed: \today}
}

@manual{openai2025,
  title        = {Function Calling},
  author       = {{OpenAI}},
  organization = {OpenAI},
  year         = {2025},
  url          = {https://platform.openai.com/docs/guides/function-calling},
  note         = {Accessed: 2026-01-27},
}

@article{chen2021evaluating,
  title={Evaluating Large Language Models Trained on Code},
  author={Chen, Mark and Tworek, Jerry and Jun, Heewoo and Yuan, Qiming and Pinto, Henrique Ponde de Oliveira and Kaplan, Jared and Edwards, Harri and Burda, Yuri and Joseph, Nicholas and Brockman, Greg and others},
  journal={arXiv preprint arXiv:2107.03374},
  year={2021},
  url={https://arxiv.org/abs/2107.03374}
}

@article{dong2024selfcollaboration,
author = {Dong, Yihong and Jiang, Xue and Jin, Zhi and Li, Ge},
title = {Self-Collaboration Code Generation via ChatGPT},
journal = {ACM Transactions on Software Engineering and Methodology},
volume = {33},
number = {7},
articleno = {189},
pages = {1--38},
year = {2024},
month = {September},
doi = {10.1145/3672459},
url = {https://doi.org/10.1145/3672459}
}

@article{wu2023autogen,
author = {Wu, Qingyun and Bansal, Gagan and Zhang, Jieyu and Wu, Yiran and Li, Beibin and Zhu, Erkang and Jiang, Li and Zhang, Xiaoyun and Zhang, Shaokun and Liu, Jiale and Awadallah, Ahmed Hassan and White, Ryen W. and Burger, Doug and Wang, Chi},
title = {AutoGen: Enabling Next-Gen LLM Applications via Multi-Agent Conversation},
journal = {arXiv preprint arXiv:2308.08155},
year = {2023},
url = {https://arxiv.org/abs/2308.08155}
}

@article{huang2024agentcoder,
author = {Huang, Dong and Zhang, Jie M. and Luck, Michael and Bu, Qingwen and Qing, Yuhao and Cui, Heming},
title = {AgentCoder: Multi-Agent-based Code Generation with Iterative Testing and Optimisation},
journal = {arXiv preprint arXiv:2312.13010},
year = {2024},
url = {https://arxiv.org/abs/2312.13010}
}

@article{austin2021program,
  title={Program Synthesis with Large Language Models},
  author={Austin, Jacob and Odena, Augustus and Nye, Maxwell and Bosma, Maarten and Michalewski, Henryk and Dohan, David and Jiang, Ellen and Cai, Carrie and Terry, Michael and Le, Quoc and others},
  journal={arXiv preprint arXiv:2108.07732},
  year={2021},
  url={https://arxiv.org/abs/2108.07732}
}

@article{yin2022natural,
  title={Natural Language to Code Generation in Interactive Data Science Notebooks},
  author={Yin, Pengcheng and Li, Wen-Ding and Xiao, Kensen and Rao, Abhishek and Wen, Yeming and Shi, Kensen and Howland, Joshua and Bailey, Paige and Catasta, Michele and Michalewski, Henryk and Polozov, Alex and Sutton, Charles},
  journal={arXiv preprint arXiv:2212.09248},
  year={2022}
}

@book{beck2002test,
  title={Test Driven Development: By Example},
  author={Beck, Kent},
  year={2002},
  publisher={Addison-Wesley Professional},
  isbn={978-0-321-14653-3}
}

@misc{copilot,
  author = {{GitHub}},
  title = {{GitHub Copilot: Your AI pair programmer}},
  year = {2021},
  url = {https://github.com/features/copilot}
}

@article{weyns2023vision,
  title={The vision of self-evolving computing systems},
  author={Weyns, Danny and B{\"a}ck, Thomas and Vidal, Rene and Yao, Xin and Belbachir, Ahmed Nabil},
  journal={Journal of Integrated Design and Process Science},
  volume={26},
  number={3-4},
  pages={351--367},
  year={2023},
  publisher={SAGE Publications Sage UK: London, England}
}

@article{jin2025reveal,
  title={ReVeal: Self-Evolving Code Agents via Iterative Generation-Verification},
  author={Jin, Yiyang and Xu, Kunzhao and Li, Hang and Han, Xueting and Zhou, Yanmin and Li, Cheng and Bai, Jing},
  volume={X},
  number={Y},
  pages={xx--yy},
  journal={arXiv preprint arXiv:2506.11442},
  year={2025}
}

@inproceedings{guo2024exploring,
  title={Exploring the potential of chatgpt in automated code refinement: An empirical study},
  author={Guo, Qi and Cao, Junming and Xie, Xiaofei and Liu, Shangqing and Li, Xiaohong and Chen, Bihuan and Peng, Xin},
  booktitle={Proceedings of the 46th IEEE/ACM International Conference on Software Engineering},
  pages={1--13},
  publisher={IEEE},
  address={Piscataway, NJ, USA},
  year={2024}
}

@inproceedings{dellarocas1998architecture,
  title={An architecture for constructing self-evolving software systems},
  author={Dellarocas, Chrysanthos and Klein, Mark and Shrobe, Howard},
  booktitle={Proceedings of the third international workshop on Software architecture},
  pages={29--32},
  publisher={ACM},
  address={New York, NY, USA},
  year={1998}
}

@article{demsar2006,
author = {Demsar, Janez},
year = {2006},
month = {01},
pages = {1-30},
title = {Statistical Comparisons of Classifiers over Multiple Data Sets},
volume = {7},
journal = {Journal of Machine Learning Research}
}

@inproceedings{Femmer2014RequirementsSmells,
  author    = {Femmer, Henning and Fernandez, Daniel Mendez and Juergens, Elmar and Klose, Michael and Zimmer, Ilona and Zimmer, J\"{o}rg},
  title     = {Rapid requirements checks with requirements smells: two case studies},
  booktitle = {Proceedings of the 1st International Workshop on Rapid Continuous Software Engineering (RCoSE '14)},
  year      = {2014},
  month     = jun,
  pages     = {10--19},
  publisher = {Association for Computing Machinery},
  address   = {New York, NY, USA},
  location  = {Hyderabad, India},
  doi       = {10.1145/2593812.2593817},
  url       = {https://doi.org/10.1145/2593812.2593817},
  isbn      = {9781450328562},
  series    = {ICSE '14}
}

@article{wu2024evolutionary,
  title={Evolutionary computation in the era of large language model: Survey and roadmap},
  author={Wu, Xingyu and Wu, Sheng-hao and Wu, Jibin and Feng, Liang and Tan, Kay Chen},
  journal={IEEE Transactions on Evolutionary Computation},
  year={2024},
  volume={XX},
  number={YY},
  publisher={IEEE},
  pages={xx--yy}
}

@misc{fang2025comprehensivesurveyselfevolvingai,
      title={A Comprehensive Survey of Self-Evolving AI Agents: A New Paradigm Bridging Foundation Models and Lifelong Agentic Systems},
      author={Jinyuan Fang and Yanwen Peng and Xi Zhang and Yingxu Wang and Xinhao Yi and Guibin Zhang and Yi Xu and Bin Wu and Siwei Liu and Zihao Li and Zhaochun Ren and Nikos Aletras and Xi Wang and Han Zhou and Zaiqiao Meng},
      year={2025},
      eprint={2508.07407},
      archivePrefix={arXiv},
      primaryClass={cs.AI},
      url={https://arxiv.org/abs/2508.07407},
}

@article{jiang2024self,
  title={Self-planning code generation with large language models},
  author={Jiang, Xue and Dong, Yihong and Wang, Lecheng and Fang, Zheng and Shang, Qiwei and Li, Ge and Jin, Zhi and Jiao, Wenpin},
  journal={ACM Transactions on Software Engineering and Methodology},
  volume={33},
  number={7},
  pages={1--30},
  year={2024},
  publisher={ACM New York, NY}
}

@article{chan2023chateval,
  title={Chateval: Towards better llm-based evaluators through multi-agent debate},
  author={Chan, Chi-Min and Chen, Weize and Su, Yusheng and Yu, Jianxuan and Xue, Wei and Zhang, Shanghang and Fu, Jie and Liu, Zhiyuan},
  journal={arXiv preprint arXiv:2308.07201},
  year={2023}
}

@article{dong2025safeguarding,
  title={Safeguarding large language models: A survey},
  author={Dong, Yi and Mu, Ronghui and Zhang, Yanghao and Sun, Siqi and Zhang, Tianle and Wu, Changshun and Jin, Gaojie and Qi, Yi and Hu, Jinwei and Meng, Jie and others},
  journal={Artificial Intelligence Review},
  volume={58},
  number={12},
  pages={382},
  year={2025},
  publisher={Springer}
}

@article{mu2024clarifygpt,
  title={Clarifygpt: A framework for enhancing llm-based code generation via requirements clarification},
  author={Mu, Fangwen and Shi, Lin and Wang, Song and Yu, Zhuohao and Zhang, Binquan and Wang, ChenXue and Liu, Shichao and Wang, Qing},
  journal={Proceedings of the ACM on Software Engineering},
  volume={1},
  number={FSE},
  pages={2332--2354},
  year={2024},
  publisher={ACM New York, NY, USA}
}

@article{Malhotra2017,
  author    = {Malhotra, Ruchika and Khanna, Megha},
  title     = {An exploratory study for software change prediction in object-oriented systems using hybridized techniques},
  journal   = {Automated Software Engineering},
  volume    = {24},
  number    = {3},
  pages     = {673--717},
  year      = {2017},
  doi       = {10.1007/s10515-016-0203-0},
  publisher = {Springer}
}

@inproceedings{paradis2025much,
  title={How much does AI impact development speed? An enterprise-based randomized controlled trial},
  author={Paradis, Elise and Grey, Kate and Madison, Quinn and Nam, Daye and Macvean, Andrew and Meimand, Vahid and Zhang, Nan and Ferrari-Church, Ben and Chandra, Satish},
  booktitle={2025 IEEE/ACM 47th International Conference on Software Engineering: Software Engineering in Practice (ICSE-SEIP)},
  pages={618--629},
  address={Piscataway, NJ, USA},
  year={2025},
  organization={IEEE},
  publisher={IEEE}
}

@article{dong2025survey,
  title={A survey on code generation with llm-based agents},
  author={Dong, Yihong and Jiang, Xue and Qian, Jiaru and Wang, Tian and Zhang, Kechi and Jin, Zhi and Li, Ge},
  journal={arXiv preprint arXiv:2508.00083},
  year={2025},
  volume={xx},
  number={yy},
  pages={1--10}
}

@inproceedings{du2024classeval,
  title     = {Evaluating Large Language Models in Class-Level Code Generation},
  author    = {Du, Xueying and Liu, Mingwei and Wang, Kaixin and Wang, Hanlin
               and Liu, Junwei and Chen, Yixuan and Feng, Jiayi and
               Sha, Chaofeng and Peng, Xin and Lou, Yiling},
  booktitle = {Proceedings of the 46th IEEE/ACM International Conference
               on Software Engineering (ICSE '24)},
  year      = {2024},
  publisher = {ACM},
  doi       = {10.1145/3597503.3639219}
}

@article{zhao2023survey,
  title={A Survey of Large Language Models},
  author={Zhao, Wayne Xin and Zhou, Kun and Li, Junyi and Tang, Tianyi and Wang, Xiaolei and Hou, Yupeng and Min, Yingqian and Zhang, Beichen and Zhang, Junjie and Dong, Zican and Du, Yifan and Yang, Chen and Chen, Yushuo and Chen, Zhipeng and Jiang, Jinhao and Ren, Ruiyang and Li, Yifan and Tang, Xinyu and Liu, Zikang and Liu, Peiyu and Nie, Jian-Yun and Wen, Ji-Rong},
  journal={arXiv preprint arXiv:2303.18223},
  year={2023},
  url={http://arxiv.org/abs/2303.18223}
}

@inproceedings{wang2021codet5,
  author = {Wang, Yue and Wang, Weishi and Joty, Shafiq and Hoi, Steven C.H.},
  title = {{CodeT5: Identifier-aware unified pre-trained encoder-decoder models for code understanding and generation}},
  booktitle = {Proceedings of the 2021 Conference on Empirical Methods in Natural Language Processing (EMNLP)},
  pages = {8696--8708},
  year = {2021},
  address = {Online and Punta Cana, Dominican Republic},
  publisher = {Association for Computational Linguistics}
}

@article{chen2021codex,
  author = {Chen, Mark and Tworek, Jerry and Jun, Heewoo and Yuan, Qiming and Pinto, Henrique Ponde de Oliveira and Kaplan, Jared and Edwards, Harri and Burda, Yuri and Joseph, Nicholas and Brockman, Greg and others},
  title = {Evaluating large language models trained on code},
  journal = {arXiv preprint arXiv:2107.03374},
  year = {2021}
}

@article{chen2023selfdebug,
  title={Teaching Large Language Models to Self-Debug},
  author={Chen, Xinyun and Lin, Maxwell and Sch{\"a}rli, Nathanael and Zhou, Denny},
  journal={arXiv preprint arXiv:2304.05128},
  year={2023}
}

@article{madaan2023selfrefine,
  title={Self-Refine: Iterative Refinement with Self-Feedback},
  author={Madaan, Aman and Tandon, Niket and Gupta, Prakhar and others},
  journal={arXiv preprint arXiv:2303.17651},
  year={2023}
}

@article{hong2023metagpt,
  author = {Hong, Sirui and Zhuge, Mingchen and Chen, Jonathan and Xiong, Xiawu and Fu, Yuheng and Cheng, Zili and Zhang, Shengyi and Wang, Jing and Zheng, Jinlin and Li, Shuyang and others},
  title = {{MetaGPT: Meta programming for multi-agent collaborative framework}},
  journal = {arXiv preprint arXiv:2308.00352},
  year = {2023},
  volume = {1},
  number = {1},
  pages = {xx--yy},
}

@inproceedings{ticoder,
  author = {Fakhoury, Sarah and Chakraborty, Saikat and Musuvathi, Madanlal and Lahiri, Shuvendu K.},
  title = {Test-driven interactive code generation},
  booktitle = {Proceedings of the 46th International Conference on Software Engineering (ICSE '24)},
  publisher = {ACM},
  address = {New York, NY, USA},
  year = {2024},
  pages = {xx--yy}
}

@misc{openai2024prompting,
author = {{OpenAI}},
title = {GPT-4.1 Prompting Guide},
year = {2024},
howpublished = {\url{https://cookbook.openai.com/examples/gpt4-1_prompting_guide}},
note = {Accessed: October 27, 2025}
}

@article{mape-k,
  author = {Kephart, Jeffrey O. and Chess, David M.},
  title = {The vision of autonomic computing},
  journal = {Computer},
  volume = {36},
  number = {1},
  pages = {41--50},
  year = {2003}
}
\end{document}